# Lorentz Boosts and Governing Equations of the Quantum Geometric Tensor in 4D Energy-Momentum Space

Yiming Pan

*State Key Laboratory of Quantum Functional Materials, School of Physical Science and Technology and Center for Transformative Science, ShanghaiTech University, Shanghai 200031, China*

## Abstract

In solids, the quantum geometric tensor (QGT), $Q_{\mu\nu} = g_{\mu\nu} - \frac{i}{2}\Omega_{\mu\nu}$, combines the quantum metric and Berry curvature. Its imaginary part governs anomalous velocity and quantized Hall transport, while its real part determines the quantum weight and geometric contributions to flat-band superfluid stiffness and dielectric response. Here, we extend the QGT to four-dimensional (4D) energy-momentum space $(\omega/c, \mathbf{k})$, treating frequency and momentum as independent coordinates. Invariance of the real-spacetime plane-wave phase fixes their Lorentz transformation in energy-momentum space. Lorentz boosts of static Dirac and Weyl systems then generate frequency metrics, quadrupolar frequency-momentum shear metrics, and electric-like Berry-Maxwell fields. We identify five Lorentz-invariant contractions and derive the governing field equations and flow conservation laws for both the quantum metric and Berry curvature from static Weyl geometry and Lorentz covariance. We further connect Gaussian spectral amplitudes of the QGT to localized Gaussian events in 4D spacetime. Finally, we discuss experimental realizations in spacetime crystals or time-varying media and clarify how this unified quantum geometry of a gravitational-like metric and electromagnetic-like gauge fields connects geometric self-consistency with physical event dynamics.

Einstein and Schrödinger sought to unify gravity and electromagnetism through Hermitian and nonsymmetric extensions of spacetime geometry [1,2]. Their construction combines the symmetric spacetime metric $g_{\mu\nu}$ and the antisymmetric electromagnetic field tensor $F_{\mu\nu}$ in a complex Hermitian tensor, $\hat{g}_{\mu\nu} = g_{\mu\nu} + i\alpha F_{\mu\nu}$, and motivates this work. In condensed-matter and photonic systems, quantum mechanics provides an intrinsic Hermitian tensor in momentum space. We obtain the gauge-invariant quantum geometric tensor (QGT) by projecting the derivatives of a quantum state $|u\rangle$ onto its orthogonal complement $\hat{P}_\perp = 1 - |u\rangle\langle u|$ [3,4]:

$$\mathcal{Q}_{\mu\nu} = \langle \partial_\mu u | (1 - |u\rangle\langle u|) | \partial_\nu u \rangle = g_{\mu\nu} - \frac{i}{2}\Omega_{\mu\nu}, \tag{1}$$

where the symmetric real part $g_{\mu\nu} = \mathrm{Re}\,\mathcal{Q}_{\mu\nu}$ defines the quantum metric and the antisymmetric imaginary part $\Omega_{\mu\nu} = -2\,\mathrm{Im}\,\mathcal{Q}_{\mu\nu}$ gives the Berry curvature. The same state $|u\rangle$ determines both the real and imaginary parts of $\mathcal{Q}_{\mu\nu}$.

In static bands, the two QGT components govern distinct physical observables. For a gapped two-dimensional (2D) Dirac cone, integrating the Berry curvature $\Omega_{xy}$ over momentum space yields a half-integer Berry flux $C = \pm 1/2$. In a lattice system, the global occupied bands and their ultraviolet (UV) completion determine the measurable Hall conductance [5,6]. Meanwhile, the real part (quantum metric $g_{ij}$) measures the quantum distance between neighboring Bloch states and sets the quantum geometric weight [7–11]. This geometric weight contributes to the superfluid stiffness in flat-band superconductors [7] and to dielectric and optical responses in insulators [9–11]. Around a gapped Dirac point, integrating the metric trace $\mathrm{Tr}(g)$ gives a logarithmic UV divergence, $\int_{|\mathbf{k}|<\Lambda_{\mathrm{UV}}} \frac{d^2k}{2\pi} \mathrm{Tr}g \simeq \frac{1}{8}\ln(\Lambda_{\mathrm{UV}}^2/m^2) + \frac{1}{8}$, while the metric and curvature obey the local bound $\mathrm{Tr}(g) \geq |\Omega_{xy}|$ [8].

Spacetime crystals, time-varying media, and spatiotemporally modulated metamaterials provide settings for periodic spatial and temporal modulation [12]. Many dynamical descriptions formulate quantum geometry in a mixed parameter space of three-dimensional (3D) momentum $\mathbf{k}$ and one-dimensional (1D) external time $t$. Such parameter–time descriptions connect the QGT to postquench wavepacket spreading [13]. They also organize the Berry connections of time-dependent states into Berry–Maxwell equations [14]. Energy-momentum Berry curvature provides a complementary description [15,16]. Building on the Berry–Maxwell formulation in Ref. [16], we extend this description to the full QGT, including its symmetric metric sector. Here, we consider normalized spectral states $|u(\mathbf{k}, \omega)\rangle$ with frequency and momentum as independent coordinates in four-dimensional (4D) energy-momentum space $k^\mu = (\omega/c, \mathbf{k})$ (Fig. 1). Because the plane-wave phase $\phi = \mathbf{k} \cdot \mathbf{r} - \omega t = \eta_{\mu\nu} k^\mu x^\nu$ is a Lorentz scalar under the Minkowski metric $\eta_{\mu\nu} = \mathrm{diag}(-1,1,1,1)$, the 4D energy-momentum coordinates $(\omega/c, \mathbf{k})$ inherit the same Lorentz transformation as real spacetime $(ct, \mathbf{r})$ [16]. For our Dirac and Weyl models, we evaluate the rest-frame eigenstates at Lorentz-boosted momenta to make their frequency dependence explicit (see End Matter).

In this *Letter*, we establish the Lorentz transformation laws, relativistic invariants, and governing field equations of the 4D QGT defined by spectral states $|u(\mathbf{k},\omega)\rangle$ in 4D energy-momentum space (Fig. 1). First, we identify five Lorentz-invariant contractions ($I_1$–$I_5$) of the QGT and apply Lorentz boosts to 1D, 2D, and 3D Dirac and Weyl systems, showing how static spatial metrics and magnetic-like Berry curvatures generate pure frequency metrics $g_{\omega\omega}$, quadrupolar frequency-momentum shear metrics $g_{\omega i}$, and electric-like Berry-Maxwell fields $\boldsymbol{\Upsilon}$. Second, we use static Weyl monopole geometry and Lorentz covariance to derive the governing field equations and flow conservation laws for both the quantum metric and Berry curvature. Third, we Fourier-transform an independent Gaussian scalar spectral function in $(\omega/c,\mathbf{k})$ space to obtain a localized Gaussian event in real 4D spacetime $(ct,\mathbf{r})$. We distinguish this localization example from physical event wavepackets. Related work demonstrates topological spacetime localization and studies nonlinear event formation and vortices [17–20].

***4D QGT and Lorentz invariants.*** As shown in Fig. 1(a), we describe a spatiotemporally modulated matter wave by normalized Floquet-Bloch states $|u(\mathbf{k},\omega)\rangle$ in 4D energy-momentum space $k^\mu = (\omega, k_x, k_y, k_z)$ (setting the light speed $c = 1$). For the models below, we evaluate the rest-frame Dirac or Weyl eigenstates at Lorentz-boosted momenta to obtain these states (see End Matter). Using the derivatives $\partial_\mu \equiv \partial/\partial k^\mu = (\partial_\omega, \nabla_\mathbf{k})$ in Eq. (1), we obtain the $4\times 4$ Hermitian QGT, $\mathcal{Q}_{\mu\nu}(\mathbf{k},\omega) = g_{\mu\nu}(\mathbf{k},\omega) - \frac{i}{2}\Omega_{\mu\nu}(\mathbf{k},\omega)$, with two coupled geometric sectors [Fig. 1(a)]. Its antisymmetric imaginary part $\Omega_{\mu\nu} = \partial_\mu\mathcal{A}_\nu - \partial_\nu\mathcal{A}_\mu$ [with 4D Berry connection $\mathcal{A}_\mu = i\langle u|\,\partial_\mu u\rangle = (-\chi,\mathcal{A})$] forms the 6-component Berry-Maxwell field tensor. We decompose it into the magnetic-like Berry curvature $\Omega_i = \frac{1}{2}\epsilon_{ijk}\Omega_{jk} = (\nabla_\mathbf{k}\times\mathcal{A})_i$ and the electric-like Berry curvature $\Upsilon_i = \Omega_{i\omega} = -\Omega_{\omega i} = -\partial_\omega A_i - \partial_i\chi$ [16]. The symmetric real part $g_{\mu\nu} = \mathrm{Re}\langle\partial_\mu u|(1-|u\rangle\langle u|)|\,\partial_\nu u\rangle$ defines the 10-component 4D quantum metric, comprising the 3D momentum metric $g_{ij}$, the pure frequency metric $g_{\omega\omega}$, and the frequency-momentum cross metric $\mathbf{g}_\omega = (g_{\omega x}, g_{\omega y}, g_{\omega z})$. Physically, the frequency metric measures how rapidly the normalized state $u(\mathbf{k},\omega)$ changes with frequency at fixed momentum. The cross metric gives the mixed contribution to quantum-state distance when frequency and momentum vary together. We use $\mathrm{Tr}g \equiv g_{xx} + g_{yy} + g_{zz}$ for the trace of the momentum block and $\mathrm{Tr}(g^2)$ for the trace of its square. We raise four-dimensional indices with $\eta^{\mu\nu}$ and choose $\varepsilon^{0123} = +1$.

Because the plane-wave phase $\phi = \mathbf{k}\cdot\mathbf{r} - \omega t = \eta_{\mu\nu}k^\mu x^\nu$ is a Lorentz scalar, a Lorentz boost in real spacetime with velocity $\boldsymbol{\beta} = \mathbf{v}/c$ induces the same coordinate transformation $\Lambda(\beta)$ in 4D energy-momentum space [Fig. 1(c)]. We obtain the boosted field components from the invariance of the quantum distance $ds^2 = 1 - |\langle u(k)|u(k+dk)\rangle|^2 = g_{\mu\nu}dk^\mu dk^\nu$ and the Berry curvature two-form $\Omega = \frac{1}{2}\Omega_{\mu\nu}dk^\mu \wedge dk^\nu$. Consider a system that is static in its rest frame $S'$ ($\partial_{\omega'}|u\rangle = 0$, so that $g'_{\omega\omega} = 0$, $\mathbf{g}'_\omega = 0$, and $\boldsymbol{\Upsilon}' = 0$). Under a Lorentz boost along the $x$-axis to the laboratory frame $S$, the energy-momentum coordinates and differentials transform as $k'_x = \gamma(k_x - \beta\omega)$, $dk'_x = \gamma(dk_x - \beta d\omega)$, $dk'_y = dk_y$, and $dk'_z = dk_z$, where $\gamma = 1/\sqrt{1-\beta^2}$ is the Lorentz factor. We substitute $dk'_x$ into the rest-frame line element $ds^2_{\mathrm{FS}} = g'_{ij}dk'_i dk'_j$ and two-form $\Omega = \frac{1}{2}\epsilon_{ijk}\Omega'_k dk'_i \wedge dk'_j$, collect terms in $d\omega$ and $d\mathbf{k}$, and obtain the boosted 4D quantum metric and Berry-Maxwell fields in frame $S$:

$$g_{\omega\omega} = \beta^2\gamma^2 g'_{xx}, \quad g_{\omega x} = -\beta\gamma^2 g'_{xx}, \quad g_{\omega a} = -\beta\gamma g'_{xa}, \tag{2}$$

$$g_{xx} = \gamma^2 g'_{xx}, \quad g_{xa} = \gamma g'_{xa}, \quad g_{ab} = g'_{ab} \quad (a, b \in y, z), \tag{3}$$

$$\Omega_x = \Omega'_x, \quad \mathbf{\Omega}_\perp = \gamma\mathbf{\Omega}'_\perp, \quad \mathbf{\Upsilon} = -\boldsymbol{\beta} \times \mathbf{\Omega} = -\beta\gamma\big(0, -\Omega'_z, \Omega'_y\big). \tag{4}$$

Equations (2)–(4) show different Lorentz scaling at corresponding spectral points: transverse Berry curvatures scale linearly as $\gamma$, while the longitudinal metric $g_{xx}$, pure frequency metric $g_{\omega\omega}$, and cross metric $g_{\omega x}$ acquire quadratic $\gamma^2$ factors.

Despite the frame dependence and $\gamma^2$ enhancement of individual components, we contract $Q_{\mu\nu}$ with the Minkowski metric $\eta^{\mu\nu} = \mathrm{diag}(-1,1,1,1)$ to obtain five Lorentz-invariant quantities:

$$\begin{aligned} I_1 &\equiv \eta^{\mu\nu} g_{\mu\nu} = \mathrm{Tr}(\mathbf{g}) - g_{\omega\omega} = \mathrm{Tr}(\mathbf{g}'), \\ I_2 &\equiv \frac{1}{2}\Omega_{\mu\nu}\Omega^{\mu\nu} = \mathbf{\Omega}^2 - \mathbf{\Upsilon}^2 = \mathbf{\Omega}'^2, \\ I_3 &\equiv -\frac{1}{8}\varepsilon^{\mu\nu\rho\sigma}\Omega_{\mu\nu}\Omega_{\rho\sigma} = \mathbf{\Omega}\cdot\mathbf{\Upsilon}, \end{aligned} \tag{5}$$

$$\begin{aligned} I_4 &\equiv g_{\mu\nu}g^{\mu\nu} = g^2_{\omega\omega} - 2\mathbf{g}^2_\omega + \mathrm{Tr}(g^2), \\ I_5 &\equiv Q_{\mu\nu}Q^{\mu\nu} = I_4 - \frac{1}{2}I_2. \end{aligned} \tag{6}$$

These invariants have direct physical interpretations: (i) $I_1$ is the Minkowski metric trace, showing how the $\gamma^2$ factor in $g_{xx}$ combines with the $-\beta^2\gamma^2$ contribution of the emergent frequency metric $g_{\omega\omega}$ to preserve an invariant 4D quantum geometric spread; (ii) $I_2$ is the Berry-Maxwell field Lagrangian and remains nonnegative for a subluminal boost of a static band structure ($|\boldsymbol{\beta}| < 1$); (iii) $I_3 = \mathbf{\Omega}\cdot\mathbf{\Upsilon}$ is a pseudoscalar density dual to the chiral anomaly ($\mathbf{E}\cdot\mathbf{B}$). For a boost from a static frame ($\mathbf{\Upsilon}' = 0$), $\mathbf{\Upsilon} = -\boldsymbol{\beta}\times\mathbf{\Omega}$ is perpendicular to $\mathbf{\Omega}$, so $I_3 = 0$ identically; a nonzero value therefore rules out a purely static rest-frame origin and signals intrinsic 4D spatiotemporal coupling. Finally, Eq. (6) relates the quadratic contractions $I_4$ and $I_5$ of the metric and 4D QGT. The five quantities are therefore not algebraically independent. For a static band structure and its Lorentz boosts, QGT positivity relates the first two invariants through the quantum geometric bound:

$$I_1 = \mathrm{Tr}(\mathbf{g}) - g_{\omega\omega} \geq \sqrt{I_2} = \sqrt{\mathbf{\Omega}^2 - \mathbf{\Upsilon}^2}, \tag{7}$$

where equality holds for the isotropic Weyl model (along with $I_4 = \frac{1}{2}I_1^2 = \frac{1}{2}I_2$ and $I_5 = 0$) but need not hold for other two-band systems. Saturation requires equal nonzero transverse metric eigenvalues; the massive 1D and 2D Dirac examples below generally give a strict inequality.

***Lorentz boost of Dirac and Weyl systems.*** We illustrate Eqs. (2)-(7) with two-band Dirac and Weyl Hamiltonians $H(\mathbf{k}') = \mathbf{d}(\mathbf{k}') \cdot \boldsymbol{\sigma}$ in 1D, 2D, and 3D (Fig. 2). For the upper band, we define the Bloch unit vector $\hat{\mathbf{n}}(\mathbf{k}') = \mathbf{d}/|\mathbf{d}|$ and write the rest-frame metric and curvature as $g'_{ij} = \frac{1}{4}(\partial'_i\hat{\mathbf{n}}) \cdot (\partial'_j\hat{\mathbf{n}})$ and $\Omega'_{ij} = -\frac{1}{2}\hat{\mathbf{n}} \cdot (\partial'_i\hat{\mathbf{n}} \times \partial'_j\hat{\mathbf{n}})$. In 1D, we consider the massive Dirac model $H_{1\mathrm{D}}(k'_x) = k'_x\sigma_x + m\sigma_z$ with energy $E' = \sqrt{k'^2_x + m^2}$ and unit vector $\hat{\mathbf{n}} = (k'_x, 0, m)/E'$. In the rest frame $S'$, the Berry curvature vanishes ($\Omega' = 0$), leaving only a bell-shaped spatial quantum metric $g'_{xx}(k'_x) = m^2/(4E'^4)$ [dashed gray curve in Fig. 2(a)]. We apply the Lorentz boost $k'_x = \gamma(k_x - \beta\omega)$ through Eq. (2) and obtain a $2 \times 2$ frequency-momentum metric in $(\omega, k_x)$ space:

$$\begin{pmatrix} g_{\omega\omega} & g_{\omega x} \\ g_{x\omega} & g_{xx} \end{pmatrix}_{1\mathrm{D}} = \frac{\gamma^2 m^2}{4[\gamma^2(k_x - \beta\omega)^2 + m^2]^2}\begin{pmatrix} \beta^2 & -\beta \\ -\beta & 1 \end{pmatrix}. \tag{8}$$

Figure 2(a) shows that, for $\beta = 0.75$ ($\gamma^2 \approx 2.29$), a Lorentz boost sharpens the longitudinal metric $g_{xx}$ and raises its peak by a factor of $\gamma^2$ (solid blue curve), even though this 1D model has no Berry curvature. The boost also generates a positive pure frequency metric $g_{\omega\omega} = \beta^2\gamma^2 g'_{xx}$ (solid green curve) and a negative cross metric $g_{\omega x} = -\beta\gamma^2 g'_{xx}$ (purple curve), while preserving the Lorentz-invariant difference $I_1 = g_{xx} - g_{\omega\omega} = m^2/(4E'^4)$. These components quantify the state's sensitivity to frequency and joint frequency-momentum variations in the geometric description of pulse deformation.

In 2D, the massive Dirac model $H_{2\mathrm{D}}(k'_x, k'_y) = k'_x\sigma_x + k'_y\sigma_y + m\sigma_z$ with $E' = \sqrt{k'^2_x + k'^2_y + m^2}$ hosts both geometric sectors. In the static frame, the out-of-plane Berry curvature $\Omega'_z = -m/(2E'^3)$ integrates to the half-integer cone flux $\mathcal{C} = -\frac{1}{2}\mathrm{sgn}(m)$. The static metric $g'_{ij} = \frac{E'^2\delta_{ij} - k'_i k'_j}{4E'^4}$ has trace $\mathrm{Tr}(g') = (k'^2 + 2m^2)/(4E'^4)$. Integrating $\mathrm{Tr}(g')$ up to a cutoff $\Lambda_{\mathrm{UV}}$ gives the quantum geometric weight $W_\Lambda = \int_{|\mathbf{k}'|<\Lambda} \frac{d^2k'}{2\pi}\mathrm{Tr}g' \simeq \frac{1}{8}\ln(\Lambda^2/m^2) + \frac{1}{8}$. This weight exhibits a logarithmic UV divergence and obeys $W_\Lambda \geq |C_\Lambda|$, where $C_\Lambda$ is the Berry flux through the same cutoff disk (see End Matter). A Lorentz boost along $x$, with $E' = \sqrt{\gamma^2(k_x - \beta\omega)^2 + k_y^2 + m^2}$, compresses the isotropic rest-frame contours (dashed gray circle) by a factor of $1/\gamma$ along $k_x$ [solid ellipse in Fig. 2(b)]. According to Eqs. (2)–(4), the boost induces a unidirectional in-plane electric-like Berry curvature [vertical streamlines and color density in Fig. 2(b)] and a $3 \times 3$ metric in $(\omega, k_x, k_y)$ space:

$$\Upsilon_y = -\frac{\beta\gamma m}{2E'^3}, \quad g_{\omega\omega} = \frac{\beta^2\gamma^2(k_y^2 + m^2)}{4E'^4}, \quad g_{\omega y} = \frac{\beta\gamma^2(k_x - \beta\omega)k_y}{4E'^4}. \tag{9}$$

The frequency-momentum shear metric $g_{\omega y}(k_x, k_y)$ has a $d_{xy}$-type quadrupolar pattern with alternating positive (+) and negative (−) lobes across the four quadrants of the $(k_x, k_y)$ plane [Fig. 2(c)].

In 3D, the massless Weyl Hamiltonian $H_{3\mathrm{D}}(\mathbf{k}') = \mathbf{k}' \cdot \boldsymbol{\sigma}$ hosts a topological monopole at $\mathbf{k}' = 0$. In the rest frame, its radial magnetic-like Berry curvature is $\boldsymbol{\Omega}' = -\mathbf{k}'/(2k'^3)$ and its transverse spherical metric is $g'_{ij} = \left(\delta_{ij}k'^2 - k'_i k'_j\right)/(4k'^4)$. For a Lorentz boost along $x$, with $k' = \sqrt{\gamma^2(k_x - \beta\omega)^2 + k_y^2 + k_z^2}$, Eqs. (2)–(4) give the 4D Weyl geometric fields:

$$\boldsymbol{\Omega} = -\frac{\gamma\left(k_x - \beta\omega, k_y, k_z\right)}{2k'^3}, \quad \boldsymbol{\Upsilon} = -\frac{\beta\gamma\left(0, k_z, -k_y\right)}{2k'^3}, \tag{10}$$

$$g_{\omega\omega} = \frac{\beta^2\gamma^2\left(k_y^2 + k_z^2\right)}{4k'^4}, \quad \mathbf{g}_\omega = \frac{\beta\gamma^2}{4k'^4}\left(-\left(k_y^2 + k_z^2\right), (k_x - \beta\omega)k_y, (k_x - \beta\omega)k_z\right). \tag{11}$$

On the transverse $\left(k_y, k_z\right)$ slice offset from the moving Weyl node ($k_x' = \gamma(k_x - \beta\omega) = a = 0.35$), the boosted monopole generates a circulating vortex of electric-like Berry curvature $\boldsymbol{\Upsilon}$ (orange streamlines) over a ring-shaped frequency metric $g_{\omega\omega}$ (green color map) [Fig. 2(d)]. The metric vanishes at the center of this offset slice ($k_y = k_z = 0$) and peaks at $\sqrt{k_y^2 + k_z^2} = |a|$. A slice through the node intersects the monopole singularity.

***Governing equations and spectral flow conservation.*** We derive the governing differential equations of both $\Omega_{\mu\nu}$ and $g_{\mu\nu}$ [Fig. 1(d)] from the static geometry of the 3D Weyl monopole and Lorentz covariance. In the rest frame $S'$, the Weyl monopole field $\boldsymbol{\Omega}' = -\mathbf{k}'/(2k'^3)$ is curl-free ($\nabla_{\mathbf{k}'} \times \boldsymbol{\Omega}' = 0$) and satisfies Gauss's law $\nabla_{\mathbf{k}'} \cdot \boldsymbol{\Omega}' = \rho'_W = -2\pi\delta^3(\mathbf{k}')$. For the rest-frame quantum metric $g'_{ij} = \left(\delta_{ij}k'^2 - k'_i k'_j\right)/(4k'^4)$, the trace $g' = 1/(2k'^2) = |\boldsymbol{\Omega}'|$ defines a root-trace scalar potential $\varphi' = \sqrt{g'/2} = 1/(2k')$ that obeys the Poisson equation $\nabla_{\mathbf{k}'}^2 \varphi' = \rho_W'$. We define the trace-reversed quantum metric $\bar{g}'_{\mu\nu} \equiv g'_{\mu\nu} - \frac{1}{2}\eta_{\mu\nu} g'$ and differentiate $\bar{g}'_{ij} = -k'_i k'_j / (4k'^4)$ to show that its momentum divergence vanishes away from the monopole core: $\partial'_i \bar{g}'_{ij} = -\frac{1}{4}\partial'_i\left(k'_i k'_j / k'^4\right) = 0$.

We transform these static identities to the laboratory frame $S$ using $\partial_{k_x'} = \gamma(\partial_{k_x} + \beta\partial_\omega)$, while leaving the transverse derivatives unchanged, to obtain the governing equations. For the imaginary sector, we transform $\nabla_{\mathbf{k}'} \times \boldsymbol{\Omega}' = 0$ and $\nabla_{\mathbf{k}'} \cdot \boldsymbol{\Omega}' = \rho'_W$ to obtain the Berry-Maxwell equations [16]:

$$\nabla_{\mathbf{k}} \cdot \boldsymbol{\Omega} = \rho_W, \quad \nabla_{\mathbf{k}} \cdot \boldsymbol{\Upsilon} = 0, \quad \nabla_{\mathbf{k}} \times \boldsymbol{\Upsilon} + \partial_\omega \boldsymbol{\Omega} = -\mathbf{j}_W, \quad \nabla_{\mathbf{k}} \times \boldsymbol{\Omega} = \partial_\omega \boldsymbol{\Upsilon}, \tag{12}$$

where $\rho_W = \gamma\rho'_W$ and $\mathbf{j}_W = \rho_W \boldsymbol{\beta}$ satisfy the monopole current continuity equation $\partial_\omega \rho_W + \nabla_{\mathbf{k}} \cdot \mathbf{j}_W = 0$. For the real metric sector, we use the covariance of the trace-reversed divergence $\partial^\mu\left(g_{\mu\nu} - \frac{1}{2}\eta_{\mu\nu} g\right) = 0$ to obtain the 4D quantum metric flow conservation laws. Here, $g \equiv I_1 = Tr\, g - g_{\omega\omega}$. In components:

$$\nabla_{\mathbf{k}} \cdot \mathbf{g}_\omega = \frac{1}{2}\frac{\partial}{\partial\omega}(\mathrm{Tr}(\mathbf{g}) + g_{\omega\omega}), \quad \partial_i g_{ij} - \frac{1}{2}\partial_j(\mathrm{Tr}(\mathbf{g}) - g_{\omega\omega}) = \frac{\partial g_{\omega j}}{\partial\omega}. \tag{13}$$

For this isotropic Weyl model, we compare Eqs. (10) and (11) and relate the real metric components to the Berry-Maxwell fields through the exact constitutive relations $g_{\omega\omega} = \boldsymbol{\Upsilon}^2/(2g)$, $\mathbf{g}_\omega =$ $-(\boldsymbol{\Upsilon} \times \boldsymbol{\Omega})/(2g)$, and $g_{ij} = \left(\delta_{ij}\boldsymbol{\Omega}^2 - \Omega_i\Omega_j - \Upsilon_i\Upsilon_j\right)/(2g)$, where $g = \sqrt{\boldsymbol{\Omega}^2 - \boldsymbol{\Upsilon}^2}$ (see End Matter). We substitute $\mathbf{g}_\omega = -(\boldsymbol{\Upsilon} \times \boldsymbol{\Omega})/(2g)$ into the first equation of Eq. (13) and obtain a *reciprocal Poynting*-type conservation law in energy-momentum space. The frequency derivative of the quantum geometric spread $U_g = (\mathrm{Tr}g + g_{\omega\omega})/2$ balances the momentum divergence of the spread flux $\mathbf{S}_g = \boldsymbol{\Upsilon} \times \boldsymbol{\Omega}/(2g)$, so that $\partial_\omega U_g + \nabla_{\mathbf{k}} \cdot \mathbf{S}_g = 0$. Note that this flow describes quantum-state spread rather than a physical energy current.

***4D spacetime Gaussian events.*** In addition to first-order flow conservation, Lorentz covariance gives second-order field equations in energy-momentum space. Away from the monopole singularity ($\rho_W = 0, \mathbf{j}_W = 0$), we take the curl of Eq. (12) to obtain a homogeneous wave equation for the Berry-Maxwell field $\Omega_{\mu\nu} = (\boldsymbol{\Upsilon}, \boldsymbol{\Omega})$. Differentiating the Weyl metric gives a nonzero source determined by its spatial variation. With $\varphi = \sqrt{g/2} = 1/(2k')$ and $\Box_k = \nabla_{\mathbf{k}}^2 - \partial_\omega^2$ in units $c = 1$, the curvature obeys a homogeneous equation, while gradients of the scalar potential source the metric equation:

$$\Box_k \Omega_{\mu\nu} = 0, \Box_k g_{\mu\nu} = 4\partial_\mu\varphi\partial_\nu\varphi. \tag{14}$$

To illustrate QGT localization in real spacetime, we consider an independent Gaussian scalar spectral function in $(\omega, \mathbf{k})$ space. Figures 3(a,b) show a Gaussian spectral amplitude and its localized Fourier envelope. The cone in Fig. 3(a) serves as a dispersion reference; frequency and momentum remain independent spectral coordinates. We write a 4D Gaussian spectral amplitude, centered at $(\mathbf{k}_0, \omega_0)$ with momentum width $\sigma_k$, frequency width $\sigma_\omega$, and phase center $(\mathbf{r}_0, t_0)$, as

$$a(\mathbf{k}, \omega) = a_0 \exp[-\frac{|\mathbf{k} - \mathbf{k}_0|^2}{2\sigma_k^2} - \frac{(\omega - \omega_0)^2}{2\sigma_\omega^2}]\exp[-i\mathbf{r}_0 \cdot (\mathbf{k} - \mathbf{k}_0) + it_0(\omega - \omega_0)], \tag{15}$$

where $a_0$ is the spectral amplitude prefactor. The QGT characterizes the normalized internal-state family across energy-momentum space. Independently, the scalar spectral function sets the spatial and temporal localization scales of its Fourier profile. The 4D Fourier transform ($\mathcal{F}_4$) from energy-momentum space $(\mathbf{k}, \omega)$ to real 4D spacetime $(\mathbf{r}, ct)$, $\tilde{a}(\mathbf{r}, t) = \int \frac{d^3k d\omega}{(2\pi)^4} a(\mathbf{k}, \omega) e^{i(\mathbf{k}\cdot\mathbf{r} - \omega t)}$, gives an explicit Gaussian event in 4D spacetime:

$$\tilde{a}(\mathbf{r}, t) = \frac{a_0 \sigma_k^3 \sigma_\omega}{(2\pi)^2} \exp\left[-\frac{|\mathbf{r} - \mathbf{r}_0|^2}{2\sigma_r^2} - \frac{(t - t_0)^2}{2\sigma_t^2}\right] e^{i(\mathbf{k}_0\cdot\mathbf{r} - \omega_0 t)}, \tag{16}$$

where the conjugate amplitude widths $\sigma_r = 1/\sigma_k$ and $\sigma_t = 1/\sigma_\omega$ set the spatial and temporal localization scales of the envelope around the spacetime coordinate $(\mathbf{r}_0, ct_0)$. Equations (15) and (16) give an exact Fourier duality ($\mathbf{k} \leftrightarrow \mathbf{r}, \omega \leftrightarrow t$): a Gaussian spectral amplitude maps one-to-one to a localized Gaussian envelope in real 4D spacetime [Fig. 3(a,b)]. The windowed 1+1-dimensional example in Fig. 3(c,d) shows how frequency-momentum correlations stretch and tilt the joint spacetime profile. The dashed line $\tau = x$ is a light-cone reference. The finite frequency window gives a nonzero differential residual (see End Matter). We use this independent scalar example to illustrate Fourier localization without constructing a source-dependent QGT field solution or independently radiating metric and curvature pulses.

***Experimental considerations.*** Our framework and the duality of Gaussian events suggest three implications for experiments and the physical interpretation of the 4D QGT:

First, the Lorentz boost of the quantum metric $g_{\mu\nu}$ and Berry curvature $\Omega_{\mu\nu}$ [Eqs. (2)-(11) and Fig. 2] suggests two experimental routes: *physical motion* and *synthetic spacetime crystals*. In the first route, a physical crystal, photonic lattice, or Dirac medium moving at velocity $\mathbf{v}$ relative to the laboratory frame undergoes a relativistic Doppler shift and Lorentz transformation in $(\omega/c, \mathbf{k})$ space. In the second route, we can design a synthetic spacetime crystal—such as an acoustic or optical metamaterial with a traveling-wave modulation $\delta n(\mathbf{r} - \mathbf{v}_m t)$—to emulate the boosted Dirac or Weyl states, with the modulation phase velocity $\mathbf{v}_m$ setting the tunable parameter $\boldsymbol{\beta} = \mathbf{v}_m/c$. We must verify the state dependence assumed in the boost construction; a traveling modulation or cone tilt alone is insufficient. The $\gamma^2$-amplified frequency metric $g_{\omega\omega}$ [Fig. 2(a,d)], quadrupolar shear metric $g_{\omega y}$ [Fig. 2(c)], and electric-like curvature $\boldsymbol{\Upsilon}$ [Fig. 2(b,d)] provide targets for ultrafast pump-probe spectroscopy and wavepacket interferometry. Numerically tested protocols in real-spectrum pseudo-Hermitian systems extract the QGT from energy fluctuations or generalized forces, suggesting complementary readout strategies [21].

Second, the governing equations of the 4D QGT [Eqs. (12) – (14)] connect its metric and curvature components through field conservation and a common microscopic state. We therefore treat the 4D quantum metric $g_{\mu\nu}(k)$ and Berry-Maxwell field $\Omega_{\mu\nu}(k)$ as related aspects of band geometry. This complements the quantum-metric contribution to geodesic-like Bloch-electron motion at second order in an electric field [22]. The analogy with gravitational and electromagnetic fields is useful here, but the Weyl metric retains a nonlinear source even away from the monopole. Its geometric conservation law alone does not establish an autonomous radiation field.

Third, the independent Gaussian scalar example illustrates Fourier localization in real spacetime [Fig. 3(a,b)]. Related work demonstrates topological events or studies nonlinear event solitons in specific media [17,19]. We ask how the coupled metric ($g_{\mu\nu}$) and curvature ($\Omega_{\mu\nu}$) influence a physical event's trajectory and deformation. The governing equations constrain how the QGT varies across energy-momentum space. To predict a matter wave event's trajectory and deformation, we must also specify the medium's dispersion and equations of motion [12,22]. We consider Hermitian bands; non-Hermitian wavepacket dynamics can involve distinct right–right and left–right QGTs [23]. Experiments can test this connection by comparing reconstructed spectral geometry with measured pulse trajectories and shapes. Thus, the historical motivation of Einstein and Schrödinger has a quantum-geometric counterpart: the real and imaginary sectors share a single Hermitian tensor. This common origin concerns quantum-state geometry; it does not unify fundamental gravitational and electromagnetic interactions.

***Conclusion.*** We formulated the 4D quantum geometric tensor $Q_{\mu\nu} = g_{\mu\nu} - \frac{i}{2}\Omega_{\mu\nu}$ in relativistic energy-momentum space $(\omega/c, \mathbf{k})$, with frequency and momentum as independent coordinates. We identified five Lorentz-invariant contractions of the 4D QGT and showed in 1D, 2D, and 3D Dirac and Weyl systems that Lorentz boosts generate pure frequency metrics $g_{\omega\omega}$, quadrupolar shear metrics $g_{\omega i}$, and electric-like Berry curvatures $\boldsymbol{\Upsilon}$. We used static Weyl geometry and Lorentz covariance to derive the governing equations and flow conservation laws. An independent Gaussian scalar spectral function in $(\omega/c, \mathbf{k})$ space maps to a localized Gaussian event in 4D spacetime. This framework provides a common quantum-geometric description of metric and gauge fields in energy-momentum space and opens an experimental route to connect spectral geometry with event wavepacket motion and deformation.

### *Acknowledgments*

We thank the discussions with Liang Zhang, Jinze He, and Zirui Zhao. We also acknowledge the use of Google Gemini Argon-4 and OpenAI GPT-6.1 Sol for fruitful discussions and cross-checks. Y. P. acknowledges support from the NSFC (No. 2023X0201-417-03) and start-up funding from ShanghaiTech University.

Please address correspondence and requests for materials to Y.P.:
yiming.pan@shanghaitech.edu.cn.

### *Data and code availability*

We will make the code repository **publicly** available upon acceptance.

### *Conflict of interest*

The author declares no competing interests.

**Figures:**

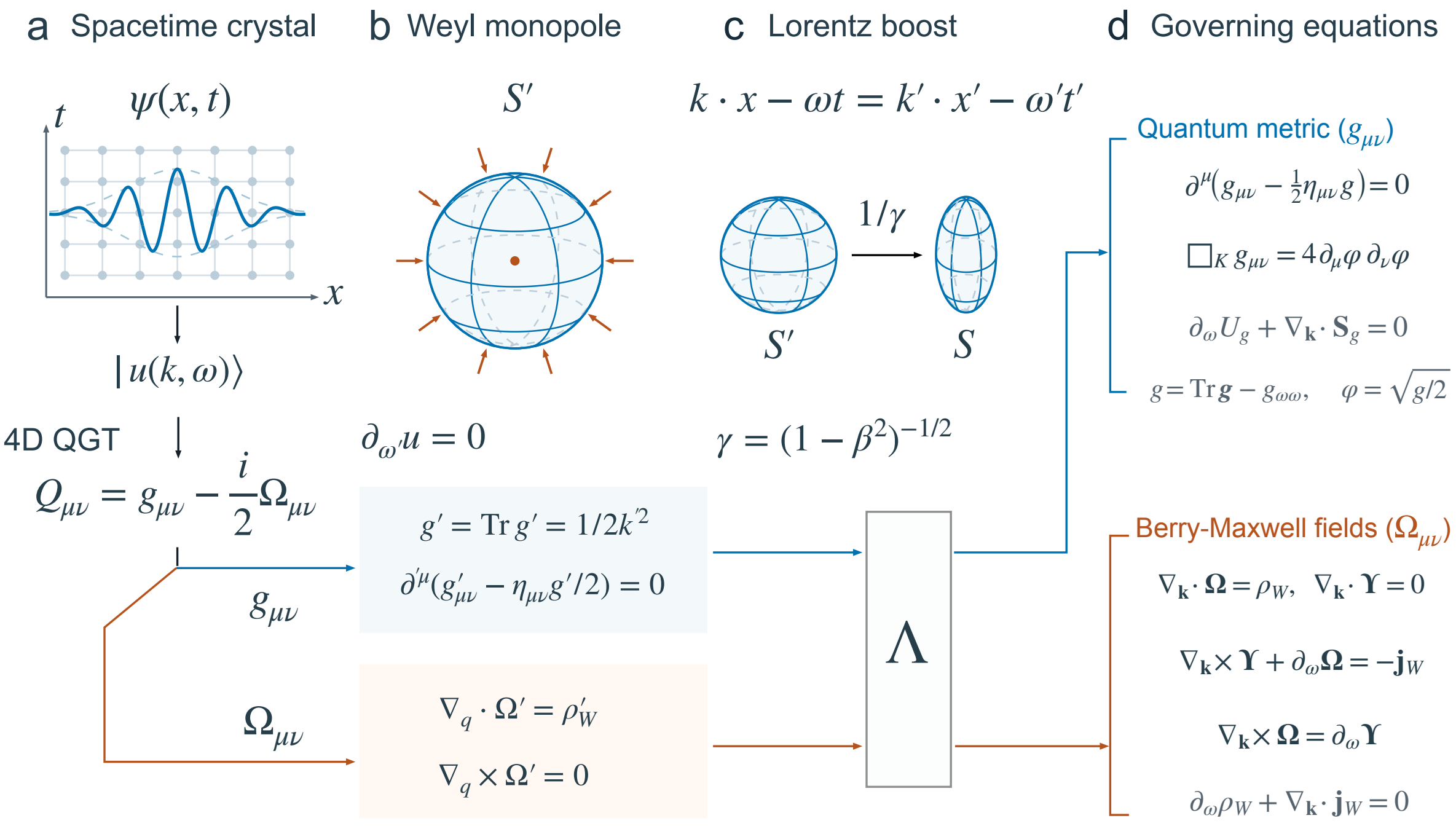

FIG. 1. From static Weyl geometry to the 4D QGT field equations. (a) A matter-wave wavepacket in a spacetime crystal defines a Floquet-Bloch state and its QGT. (b) Static Weyl geometry in the rest frame $S'$. (c) A phase-preserving boost compresses fixed-frequency momentum contours by a factor of $1/\gamma$. (d) We derive the metric and Berry–Maxwell field equations together with their flow conservation laws. Blue and orange denote the metric and curvature sectors, respectively. Here $s = \mathrm{Tr}g - g_{\omega\omega}$, $\varphi = \sqrt{s/2}$, and $c = 1$. The second-order metric equation applies away from the monopole core.

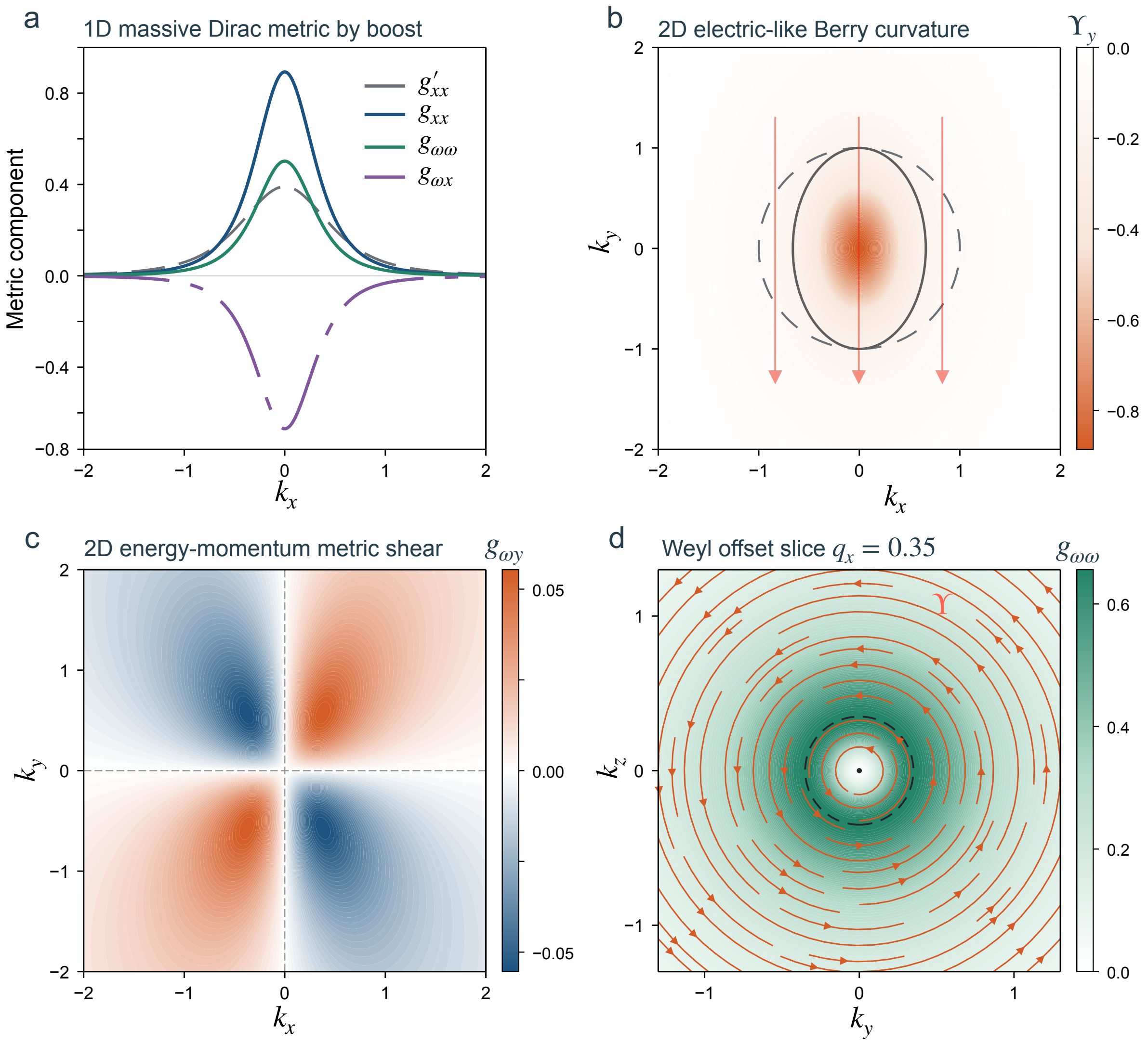

FIG. 2. Lorentz-boosted Dirac and Weyl geometry for $\beta = 0.75$, $\omega = 0$, $c = 1$, and $m = 0.8$ in the massive models. (a) Rest-frame and boosted 1D metric components. (b) Negative $\Upsilon_y$ in the 2D Dirac model; the dashed circle and solid ellipse show Lorentz compression. (c) Quadrupolar mixed metric $g_{\omega y}$. (d) Weyl frequency metric $g_{\omega\omega}$ (green) and electric-like vortex $\boldsymbol{\Upsilon}$ (orange) on the offset slice $q_x = a = 0.35$. The dashed ring $r_\perp = |a|$ marks the metric maximum; the offset slice excludes the Weyl node.

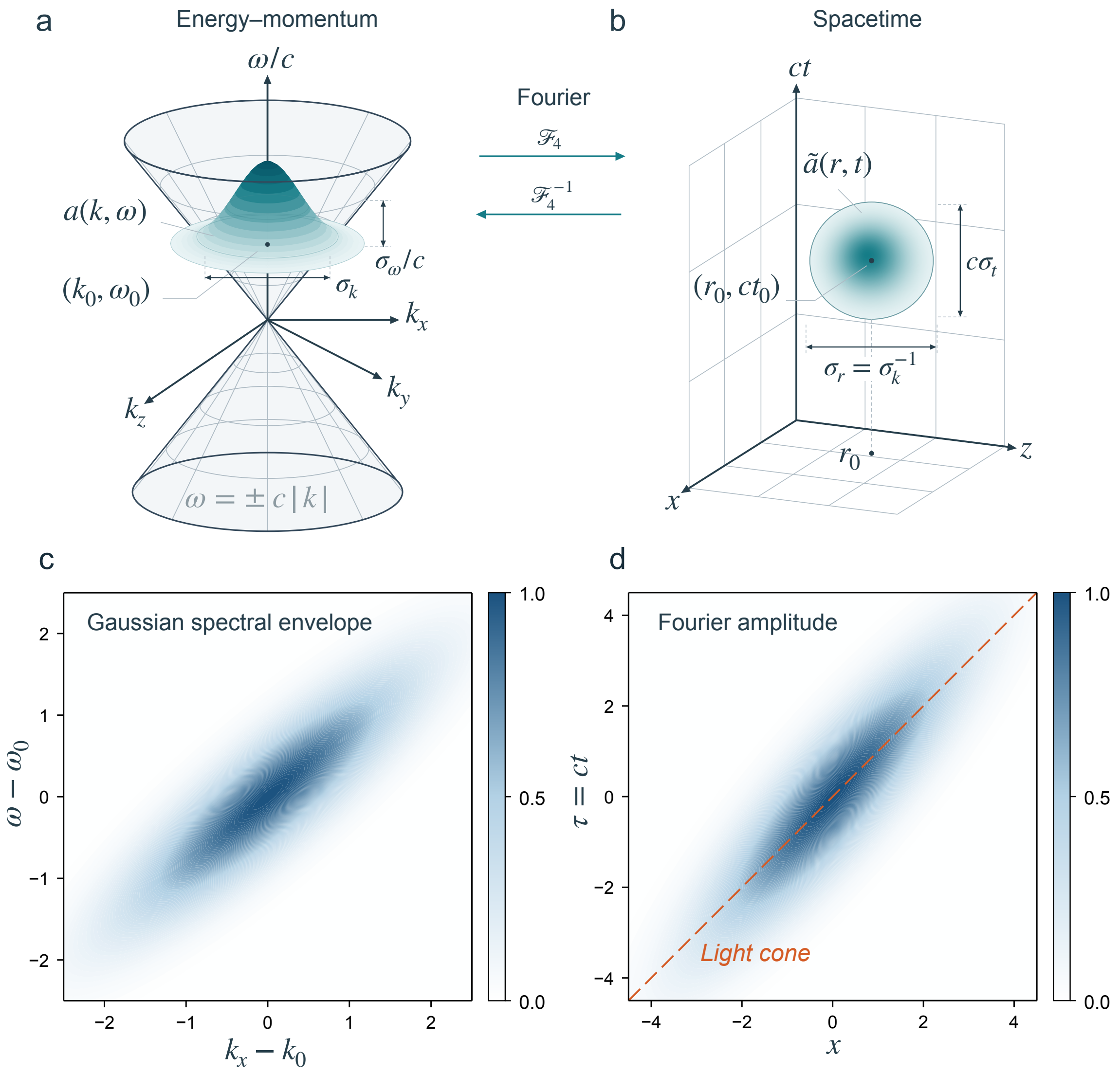

FIG. 3. Fourier duality of a Gaussian-shaped QGT and its localized spacetime profile. (a,b) An independent scalar spectral amplitude $a(\mathbf{k},\omega)$ and its spacetime envelope $\tilde{a}(\mathbf{r},t)$, with conjugate widths $\sigma_r = 1/\sigma_k$ and $\sigma_t = 1/\sigma_\omega$. The cone in (a) is a dispersion reference. (c) Normalized spectral envelope $a = f(\delta k_x - \delta\omega)W(\delta\omega)$, where $\delta k_x = k_x - k_0$ and $\delta\omega = \omega - \omega_0$; the Gaussian profile and frequency window have amplitude widths $\sigma_k = 0.55$ and $\sigma_\omega = 1$, respectively. (d) Normalized Fourier amplitude $|\tilde{a}(x,\tau)|$, with $\tau = ct$ and $c = 1$. The dashed line $\tau = x$ marks one light-cone branch.

## End Matter

***Floquet-Bloch states, Lorentz Jacobian, and geometric bounds.***

We use $c = 1$ throughout the main text and End Matter, where $c$ denotes the speed of light in vacuum. The 4D energy-momentum coordinates $k^\mu = (\omega/c, \mathbf{k})$ are conjugate to real spacetime $x^\mu = (ct, \mathbf{r})$, with Minkowski metric $\eta_{\mu\nu} = \mathrm{diag}(-1,1,1,1)$. We write $\tau = ct = t$ for the real-space Fourier plot and $\Box_k = \nabla_k^2 - \partial_\omega^2$ for the wave operator in these units. We treat $\omega$ and $\mathbf{k}$ as independent spectral coordinates and construct the laboratory-frame state $|u(\mathbf{k},\omega)\rangle$ for the Dirac and Weyl models by evaluating the static eigenstate $|u_0(\mathbf{k}')\rangle$ at the boosted momentum:

$$|u(\mathbf{k},\omega)\rangle = |u_0(\mathbf{k}')\rangle, \quad \mathbf{k}' = \big(\gamma(k_x - \beta\omega), k_y, k_z\big), \tag{A1}$$

where $\mathbf{q} \equiv \mathbf{k}' = \big(q_x, k_y, k_z\big)$ is the rest-frame momentum label used in Figs. 1 and 2(d). Equation (A1) defines the spectral family in a fixed internal basis by evaluating the rest-frame state at the boosted momentum. The chain rule $\partial_\mu = \Lambda_\mu^\alpha \, \partial'_\alpha$, with Jacobian $\Lambda_\mu^\alpha = \partial k'^\alpha / \partial k^\mu$, gives the QGT transformation $Q_{\mu\nu} = \Lambda_\mu^\alpha \Lambda_\nu^\beta Q'_{\alpha\beta}$. Since $\partial_{\omega'}|u_0\rangle = 0$ in $S'$, only spatial indices $\alpha, \beta \in \{x, y, z\}$ enter with $\Lambda_\omega^x = -\beta\gamma$, $\Lambda_x^x = \gamma$, and $\Lambda_y^y = \Lambda_z^z = 1$, yielding Eqs. (2)–(4). In Eq. (6), index order matters: $I_5 \equiv Q^{\mu\nu}Q_{\mu\nu} = I_4 - \frac{1}{2}I_2$, whereas $Q^{\mu\nu}Q_{\nu\mu} = I_4 + \frac{1}{2}I_2$. To prove Eq. (7), we align $\mathbf{\Omega}'$ along axis 3 in $S'$. Positive semi-definiteness of $Q'_{ij}$ requires $g'_{11}g'_{22} - (g'_{12})^2 \geq (\Omega'_3)^2/4$, so $\mathrm{Tr}\mathbf{g}' \geq 2\sqrt{g'_{11}g'_{22}} \geq |\mathbf{\Omega}'|$. Lorentz invariance of $I_1 = \mathrm{Tr}\mathbf{g}'$ and $I_2 = |\mathbf{\Omega}'|^2$ gives $I_1 \geq \sqrt{I_2}$ in every frame. Equality holds if and only if $g'_{11} = g'_{22} = |\mathbf{\Omega}'|/2$ and $g'_{12} = g'_{33} = 0$ (Weyl monopole).

***Explicit Lorentz-boosted forms for 1D, 2D, and 3D models.***

We substitute $dk'_x = \gamma(dk_x - \beta d\omega)$ and $\mathbf{k}'_\perp = \mathbf{k}_\perp$ into the rest-frame quantum geometric tensor of the massive Dirac and massless Weyl models to obtain the laboratory-frame field components. For the 1D massive Dirac model in the $(1+1)$-dimensional $(\omega, k_x)$ space, the Lorentz-boosted $2 \times 2$ frequency-momentum quantum metric is:

$$g_{\mu\nu}^{1\mathrm{D}} = \frac{\gamma^2 m^2}{4[\gamma^2(k_x - \beta\omega)^2 + m^2]^2}\begin{pmatrix} \beta^2 & -\beta \\ -\beta & 1 \end{pmatrix}, \quad \Omega_{\mu\nu}^{1\mathrm{D}} = 0. \tag{A2}$$

For the 2D massive Dirac model in the $(2+1)$-dimensional $\big(\omega, k_x, k_y\big)$ space, with $k'_x = \gamma(k_x - \beta\omega)$ and $E' = \sqrt{k'^2_x + k_y^2 + m^2}$, the Lorentz-boosted $3 \times 3$ quantum metric is:

$$g_{\mu\nu}^{2\mathrm{D}} = \frac{1}{4E'^4}\begin{pmatrix} \beta^2\gamma^2\big(k_y^2 + m^2\big) & -\beta\gamma^2\big(k_y^2 + m^2\big) & \beta\gamma k'_x k_y \\ -\beta\gamma^2\big(k_y^2 + m^2\big) & \gamma^2\big(k_y^2 + m^2\big) & -\gamma k'_x k_y \\ \beta\gamma k'_x k_y & -\gamma k'_x k_y & k'^2_x + m^2 \end{pmatrix}. \tag{A3}$$

The corresponding Lorentz-boosted magnetic-like curvature $\Omega_z = \Omega_{xy}$ and emergent electric-like curvature $\Upsilon_y = -\Omega_{\omega y}$ are:

$$\Omega_z = -\frac{\gamma m}{2E'^3}, \quad \Upsilon_y = -\frac{\beta\gamma m}{2E'^3}, \quad \Upsilon_x = 0. \tag{A4}$$

For the 3D massless Weyl monopole in the $(3+1)$-dimensional $(\omega, \mathbf{k})$ space, with $k' = \sqrt{k'^2_x + k_y^2 + k_z^2}$ and $k_\perp^2 = k_y^2 + k_z^2$, the full $4 \times 4$ quantum metric $g_{\mu\nu}^{3\mathrm{D}}$ is:

$$g_{\mu\nu}^{3\mathrm{D}} = \frac{1}{4k'^4}\begin{pmatrix} \beta^2\gamma^2 k_\perp^2 & -\beta\gamma^2 k_\perp^2 & \beta\gamma k'_x k_y & \beta\gamma k'_x k_z \\ -\beta\gamma^2 k_\perp^2 & \gamma^2 k_\perp^2 & -\gamma k'_x k_y & -\gamma k'_x k_z \\ \beta\gamma k'_x k_y & -\gamma k'_x k_y & k'^2_x + k_z^2 & -k_y k_z \\ \beta\gamma k'_x k_z & -\gamma k'_x k_z & -k_y k_z & k'^2_x + k_y^2 \end{pmatrix}. \tag{A5}$$

The boosted Berry-Maxwell fields $\mathbf{\Omega}$ and $\mathbf{\Upsilon} = -\boldsymbol{\beta} \times \mathbf{\Omega}$ are:

$$\mathbf{\Omega} = -\frac{\gamma}{2k'^3}\begin{pmatrix} k_x - \beta\omega \\ k_y \\ k_z \end{pmatrix}, \quad \mathbf{\Upsilon} = -\frac{\beta\gamma}{2k'^3}\begin{pmatrix} 0 \\ k_z \\ -k_y \end{pmatrix}. \tag{A6}$$

***Ultraviolet scaling and topological bound of the 2D Dirac quantum weight.***

For the 2D massive Dirac model $H_{2\mathrm{D}}(\mathbf{k}') = k'_x\sigma_x + k'_y\sigma_y + \mathrm{m}\sigma_z$ with $E' = \sqrt{k'^2 + m^2}$, we integrate the rest-frame metric trace $\mathrm{Tr}\mathbf{g}' = (k'^2 + 2m^2)/(4E'^4)$ and Berry curvature $\Omega'_z = -m/(2E'^3)$ over the UV cutoff disk $k' < \Lambda_{\mathrm{UV}}$ with measure $d^2\mathbf{k}'/(2\pi)$ to obtain:

$$W_\Lambda = \frac{1}{8}\left[\ln\left(1 + \frac{\Lambda_{\mathrm{UV}}^2}{m^2}\right) + \frac{\Lambda_{\mathrm{UV}}^2}{\Lambda_{\mathrm{UV}}^2 + m^2}\right], \tag{A7}$$

$$C_\Lambda = -\frac{\mathrm{sgn}(m)}{2}\left(1 - \frac{|m|}{\sqrt{\Lambda_{\mathrm{UV}}^2 + m^2}}\right), \; W_\Lambda \geq |C_\Lambda|. \tag{A8}$$

The pointwise inequality $\mathrm{Tr}\mathbf{g}' - |\Omega'_z| = (E' - |m|)^2/(4E'^4) \geq 0$ implies $W_\Lambda \geq |C_\Lambda|$ on any cutoff disk. For $\Lambda_{\mathrm{UV}} \gg |m|$, $C_\Lambda \to -\mathrm{sgn}(m)/2$ gives the single-cone half-integer flux. The full lattice Hall conductance depends on the global occupied bands and their UV completion [5,6]. In this limit, the quantum geometric weight $W_\Lambda \simeq [\ln(\Lambda_{\mathrm{UV}}^2/m^2) + 1]/8$ diverges logarithmically. More generally, quantum geometric weight enters the geometric contributions to flat-band superfluid stiffness and dielectric response [7,10,11].

***Weyl monopole constitutive reconstruction and governing field equations.***

For the 3D Weyl monopole $H_{3\mathrm{D}}(\mathbf{k}') = \mathbf{k}' \cdot \boldsymbol{\sigma}$, let $k' = |\mathbf{k}'| = \sqrt{\gamma^2(k_x - \beta\omega)^2 + k_y^2 + k_z^2}$. The invariant trace $g \equiv I_1 = \mathrm{Tr}\mathbf{g} - g_{\omega\omega}$ (written as $s \equiv g$ in Fig. 1) saturates Eq. (7): $g = \sqrt{\mathbf{\Omega}^2 - \mathbf{\Upsilon}^2} = 1/(2k'^2)$. We eliminate $\mathbf{k}'$ and $\boldsymbol{\beta}$ between Eqs. (10) and (11) to reconstruct all ten metric components from $(\mathbf{\Omega}, \mathbf{\Upsilon})$:

$$g_{\omega\omega} = \frac{\mathbf{\Upsilon}^2}{2\mathrm{g}}, \quad \mathbf{g}_\omega = -\frac{\mathbf{\Upsilon} \times \mathbf{\Omega}}{2\mathrm{g}}, \tag{A9}$$

$$g_{ij} = \frac{\delta_{ij}\mathbf{\Omega}^2 - \Omega_i\Omega_j - \Upsilon_i\Upsilon_j}{2\mathrm{g}}. \tag{A10}$$

In $S'$, the monopole flux is $-2\pi$, so $\nabla_{\mathbf{k}'} \cdot \mathbf{\Omega}' = \rho'_W = -2\pi\delta^3(\mathbf{k}')$. We account for the longitudinal delta-function Jacobian $\delta(k'_x) = \gamma^{-1}\delta(k_x - \beta\omega)$, so the boosted monopole density $\rho_W = \gamma\rho'_W$ and current $\mathbf{j}_W = \rho_W \boldsymbol{\beta}$ obey Eq. (12). The root-trace potential $\varphi = \sqrt{\mathrm{g}/2} = 1/(2k')$ satisfies:

$$\Box_k \varphi = \nabla^2_{\mathbf{k}'}\varphi = -2\pi\delta^3(\mathbf{k}') = \frac{\rho_W}{\gamma}. \tag{A11}$$

For the trace-reversed metric $\bar{g}_{\mu\nu} \equiv g_{\mu\nu} - \frac{1}{2}\eta_{\mu\nu}g$, we differentiate $\bar{g}'_{ij} = -k'_i k'_j/(4k'^4)$ and obtain $\partial'_i \bar{g}'_{ij} = 0$ on $\mathrm{k}' > 0$. No singular $\delta^3(\mathbf{k}')$ source arises in this divergence at $\mathbf{k}' = 0$ because the small-sphere surface integral is proportional to the vanishing odd angular average $\oint \hat{n}_j d\Omega = 0$. We boost $\partial^\mu \bar{g}_{\mu\nu} = 0$ to obtain Eq. (13) and the reciprocal Poynting law $\partial_\omega U_g + \nabla_{\mathbf{k}} \cdot \mathbf{S}_g = 0$. At second order, we evaluate the Laplacian $\nabla^2_{\mathbf{k}'} g'_{ij} = k'_i k'_j/k'^6 = 4(\partial'_i\varphi)(\partial'_j\varphi)$ on $\mathrm{k}' > 0$ and boost the result to obtain Eq. (14) and its trace:

$$\Box_k g_{\mu\nu} = 4(\partial_\mu\varphi)(\partial_\nu\varphi), \quad \Box_k \mathrm{g} = 4g^2. \tag{A12}$$

The $k'^{-4}$ singularity at $\mathbf{k}' = 0$ in the second-order source requires a core mass regularization or finite-part extension. On the regular offset slice $q_x \equiv k'_x = \mathrm{a} = 0.35$ [Fig. 2(d)], $g_{\omega\omega} = \beta^2\gamma^2 r_\perp^2/[4(a^2 + r_\perp^2)^2]$ peaks on the ring $r_\perp \equiv \sqrt{k_y^2 + k_z^2} = |a|$.

***Windowed spectral envelopes and QGT event wavepacket.***

For Figs. 3(c,d), we define $\delta k_x = k_x - k_0$ and $\delta\omega = \omega - \omega_0$, with $c = 1$. The characteristic Gaussian $f(q) = \exp[-q^2/(2\sigma_k^2)]$, where $q = \delta k_x - \delta\omega$, satisfies $(\partial_{k_x} + \partial_\omega)\mathrm{f} = 0$. We apply the frequency window $W(\delta\omega) = \exp(-\frac{(\delta\omega)^2}{2\sigma_\omega^2})$ to obtain $a = fW$ and $(\partial_{k_x} + \partial_\omega)a = -\frac{\delta\omega}{\sigma_\omega^2}a$. We Fourier-transform the windowed amplitude in these detunings, with measure $\mathrm{d}k_x d\omega/(2\pi)^2$ and $\tau = ct = t$, to obtain:

$$\tilde{a}(x,\tau) = \frac{\sigma_k\sigma_\omega}{2\pi}\exp\left[-\frac{\sigma_k^2 x^2}{2} - \frac{\sigma_\omega^2(x-\tau)^2}{2}\right]. \tag{A13}$$

Spectral windowing produces the nonzero residual above. Here, $a(\mathbf{k},\omega)$ is an independent scalar spectral function of the QGT, and its 4D Fourier transform $\tilde{a}(\mathbf{r},t)$ illustrates spacetime localization. Neither function is identified here with a physical matter-wave or optical-pulse envelope multiplying the normalized internal state $|u(\mathbf{k},\omega)\rangle$. Equations (12)–(14) constrain how the components of $Q_{\mu\nu}(\mathbf{k},\omega)$ vary across $(\mathbf{k},\omega)$ space. A physical matter-wave field $|\Psi(\mathbf{r},t)\rangle$ instead obeys the medium's dynamical equations. In semiclassical wavepacket descriptions, the curvature $\Omega_{\mu\nu}$ drives anomalous deflection, while the metric $g_{\mu\nu}$ contributes to finite-width spreading and geodesic-like gradient forces [12,22]. Figure 2 uses the upper band with $\beta = 0.75$ and $\mathrm{m} = 0.8$ for the massive models; Figs. 3(c,d) use $\sigma_k = 0.55$ and $\sigma_\omega = 1$.